\documentstyle[multicol,tighten,aps,epsfig]{revtex} 

\begin{document}
\draft

\def\coal{$A+A\rightarrow A$}
\def\annil{$A+A\rightarrow 0$}
\def\b{\beta}
\def\a{\alpha}
\def\half{{1\over2}}
\def\bk{{\b^2\over k}}
\def\NI{\noindent}
\def\qa{{\hat q}_a}
\def\qb{{\hat q}_b}
\def\z{{\overline{z}}}
\def\k{{\overline{k}}}
\def\half{{1\over2}}
\def\kR{k_{\rm R}}
\def\kI{k_{\rm I}}
\def\C{{\rm\bf C}}
\def\O{{\cal O}}

\title{Kinetics of Coalescence, Annihilation, and the q-State Potts Model
in One Dimension}

\author{Thomas Masser and Daniel ben-Avraham\footnote{{\bf e-mail:}
benavraham@clarkson.edu}}

\address{Physics Department, and Clarkson Institute for Statistical
Physics (CISP),\\
Clarkson University, Potsdam, NY 13699-5820}
\maketitle

\begin{abstract}
\NI{\bf Abstract:}
The kinetics of the q-state Potts model in the zero temperature limit in one
dimension is analyzed exactly through a generalization of the method of empty
intervals, previously used for the analysis of diffusion-limited coalescence,
\coal.  In this new approach, the q-state Potts model, coalescence, and
annihilation (\annil) all satisfy the same diffusion equation, and differ
only in the imposed initial condition.
\end{abstract}

\begin{multicols}{2}

The kinetics of the $q$-state Potts model has drawn
much recent attention, in particular regarding its persistence
probabilities~[1-10].  An understanding has emerged that, as far as
persistence is concerned, the kinetics of the $q$-state Potts model in the
zero temperature limit, with Glauber dynamics and in one dimension, is
equivalent to that of the Ising model, when a
fraction $1/q$ of the spins are initially up~[6].  It is also understood that
if one regards the boundaries between domains as particles (kinks), then $q=2$
(the Ising model) corresponds to the diffusion-limited annihilation process
\annil, while the limit $q\to\infty$ corresponds to coalescence, \coal.  Here
we extend the formalism of empty intervals, previously used for the exact
analysis of
\coal~[11], to annihilation, by considering the probability that an
interval contains an even number of particles, and
we apply this new approach also to the $q$-state Potts model.  This results in
one simple diffusion equation describing all three models (annihilation, Potts
model, and coalescence): they differ only in the initial
conditions they impose.

Consider diffusion-limited annihilation, \annil, in a one-dimensional
lattice of lattice spacing $a$.  The particles hop randomly to the nearest
site to their right or left, at equal rate $\Gamma$, and annihilate
immediately upon encounter.  Let
$G_n(t)$ be the probability that an arbitrary segment of $n$ consecutive
sites contains an even number of particles, at time $t$. (We assume that the
system is infinite and homogeneous.)  A site can be either empty or occupied
by a single particle, so the particle concentration is
$$
c(t)={1-G_1(t)\over a}\;.  \eqno(1)
$$ 

Because the reaction \annil\ conserves the number of particles modulo~2, the
only way that the parity within an $n$-segment is changed is when
particles at the edge of the segment hop out, or when particles outside
of the segment hop in.  To describe these events, we require $F_n(t)$ ---
the probability that an $n$-segment containing an even number of particles is
followed by the presence of a particle at the $(n+1)$-th site. 
This can be expressed in terms of the $G_n$ (Fig.~1):
$$ 
2F_n(t)=(1-G_1)+(G_n-G_{n+1}) \;.     \eqno(2a)
$$
Likewise, $H_n(t)$ --- the probability that an $n$-segment containing an {\it
odd\/} number of particles is followed by a particle at the $(n+1)$-th site
--- is
$$
2H_n(t)=(1-G_1)-(G_n-G_{n+1}) \;.     \eqno(2b)
$$
\begin{figure}
\centerline{\epsfxsize=4cm \epsfbox{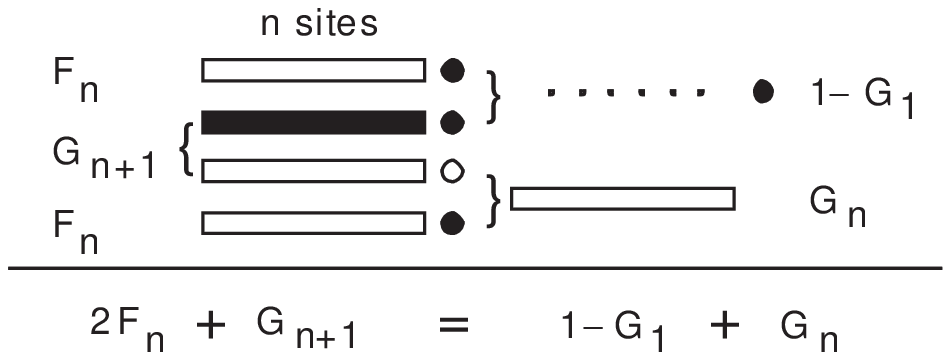}}
\smallskip
\noindent{{\bf Fig.~1:} Computation of $F_n(t)$.
Empty (solid) rectangles symbolize $n$-segments with an even (odd)
number of particles. Empty (solid) circles represent empty (occupied) sites.
The four events listed may be grouped in two different ways, yielding
eq.~(2a).  One can also see that the probabilities of the first two events
add up to $F_n+H_n=1-G_1$, yielding eq.~(2b).}
\end{figure} 
The evolution equation for $G_n$ is now readily obtained:
$$
{\partial\over\partial t}G_n(t)=
2\Gamma(F_{n-1}-H_{n-1}+H_n-F_n)\;,  
$$
where the first term on the r.h.s expresses the event that a particle at site
$n$ jumps out of the segment, leaving an even number of particles in the
remaining $n-1$ sites (and hence in the $n$-segment); the second term denotes
the same case, but when there are initially an even number of particles in
the $n$-segment (that is, an odd number in the $(n-1)$-segment); and the last
two terms pertain to a particle just outside of the $n$-segment, at site
$n+1$, jumping in.  The factor of
$2$ accounts for the fact that each event may occur at the left or right
edge of the segment, with equal probability.  Using eqs.~(2), this becomes
$$
{\partial\over\partial t}G_n(t)=
2\Gamma(G_{n-1}-2G_n+G_{n+1})\;.  \eqno(3) 
$$

The case of $n=1$ requires a special equation, since $G_0$ is undefined. 
Taking into account all the ways $G_1$ might change, one finds
$$
{\partial\over\partial t}G_1(t)=
2\Gamma(1-2G_1+G_2)\;.   
$$
Thus, eq.~(3) may be understood to be valid for all $n\geq 1$, provided that
one uses the boundary condition
$$
G_0(t)=1\;.  \eqno(4a)
$$
Additionally, since the $G_n$ are {\it probabilities\/}, we have
$$
0\leq G_n(t)\leq 1\;.  \eqno(4b)
$$

We now turn our attention to the $q$-state Potts model in one dimension,
at the zero temperature limit and with Glauber dynamics.  (This is also known
as the {\it voters model\/}, with $q$ opinions.)  Each site can assume one
of $q$ states, $s_1,\,s_2,\dots,s_q$.  The model possesses $q$ equivalent
ground states, where all the sites are in the same state.  If the system is
started away from equilibrium, then neighboring sites of differing states
react according to the scheme:
$$
s_is_j\to\cases{
  s_is_i &rate $\Gamma\;,$ \cr
  s_js_j &rate $\Gamma\;.$ \cr }  \eqno(5)
$$
Hence the interface between domains of different states may be regarded as
particles (kinks) that perform regular random walks.  Let the kinks between a
domain of state
$s_i$ on the left, and a domain of state $s_j$ on the right, be denoted by
$A_{ij}$.  The dynamics~(5) implies the following immediate reactions between
colliding kinks:
$$
A_{ij}+A_{jk}\to A_{ik}\;,  \eqno(6a)
$$
$$
A_{ij}+A_{ji}\to 0\;.  \eqno(6b)            
$$
Notice that the inner indices ($j$, in the cases above) must be the same for
kinks to meet.

Focusing now on the kinks, and assuming an infinite homogeneous system, let
$G_n(t)$ be the probability that an arbitrary $n$-segment contains kinks
such that their total number of indices {\it of each of the {\rm q} kinds} is
even, at time $t$.  Because the reactions~(6) conserve the parity of the
numbers of indices, reactions within an $n$-segment do not affect $G_n$. 
Thus, $G_n$ changes only through kinks that hop into or out of the segment,
exactly as in the case of the annihilation process.  (Note that a kink
must have different indices, so the parity is always affected by such
transitions.)  It follows that the $G_n$ obey eq.~(3), with the boundary
conditions~(4).  Finally, if one merely counts the kinks, ignoring their
indices, then their concentration is given once again by eq.~(1).

For $q=2$ the process~(6a) never takes place (it requires at least three
states) and the problem reduces to that of the annihilation process, \annil.
On the other hand, the probability that the sites to the left and right of
the two kinks are in the same state $s_i$ tends to zero as $q\to\infty$. In
that case process~(6b) never takes place and the model is then analogous to
coalescence, \coal.  For $2<q<\infty$ kinks may either coalesce or
annihilate, and we have a mixture of the two processes.
Annihilation, coalescence, and the $q$-state Potts model are all described by
the same (discrete) diffusion equation, eq.~(3), and the same boundary
conditions, eqs.~(4).

The difference between the three models is in their implied
initial condition.  Assume, for example, initial conditions where each site
is randomly in one of its $q$ possible states.  The probability that an
$n$-segment contains even numbers of indices equals the probability that
the first and last site in the segment are in the same state.  For the case of
initially random states this yields
$$
G_n(0)={1\over q}\;.  \eqno(7)
$$
While the initial state above is natural, it is by no means unique. 
Consider, for example, the initial state of Fig.~2a, for which
$$
G_n(0)=\cases{
{m-n\over m} &$0<n\leq m\;,$ \cr
1 &$kqm<n\leq(kq+1)m$,\ \ \ $k=1,2,\dots$ \cr  
0 & otherwise.}  \eqno(8)
$$
This corresponds to kinks that are initially evenly spaced, $m$ sites apart
from each other.  Notice, however, that the initial states in between the
kinks could be chosen differently (Fig.~2b), thus dictating a different
form of $G_n(0)$ and different kinetics!

\begin{figure}
\centerline{\epsfxsize=6cm \epsfbox{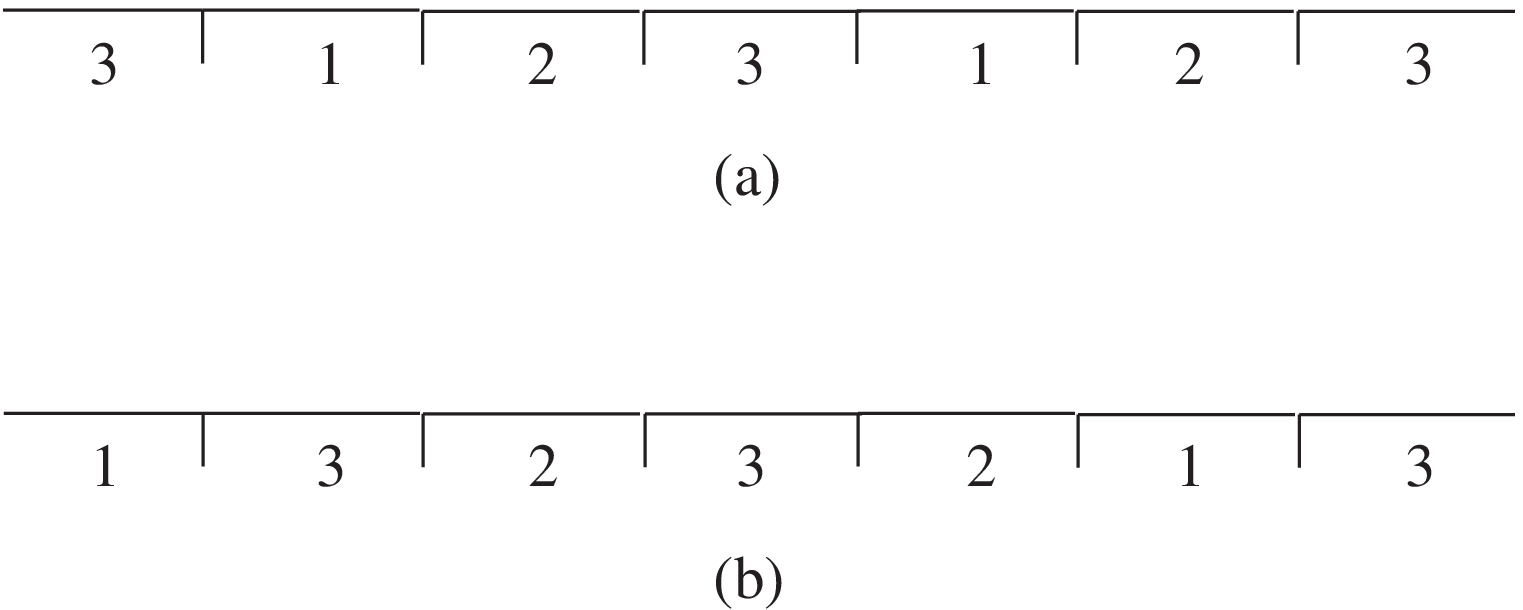}}
\noindent{{\bf Fig.~2:} Possible initial configurations for the $q$-state Potts
model.  Here the system consists of alternating domains of equal length $m$,
which corresponds to equally spaced kinks (the vertical bars), $m$ sites apart
from each other.  In (a) the states of the domains alternate periodically:
$1,2,\dots,q,1,2,\dots$ while in (b) the states are random (but do not
repeat).  In this illustration, $q=3$.}
\end{figure} 

Eq.~(3) may be solved by standard techniques, for
example, by Laplace-transforming with respect to time, fitting an exponential
solution to the resulting difference equation, and finally inverting the
Laplace-transformed solution~[12].  With the boundary conditions~(4) and the
natural initial condition~(7), one obtains:
$$
G_n(\tau)={1\over q}+{q-1\over q}\int_0^{\tau}ne^{-2\tau'}I_n(2\tau')
  \,{d\tau'\over2\tau'}\;,
$$
where $\tau=2\Gamma t$ and $I_n(\cdot)$ is the modified Bessel function of
order $n$~[13].  In particular, the probability that a site contains a kink,
is
$$
1-G_1(\tau)={q-1\over q}\Big(1-\int_0^{\tau}e^{-2\tau'}I_1(2\tau')
  \,{d\tau'\over2\tau'}\Big)\;.  \eqno(9)
$$
Thus, with natural initial conditions, the density of kinks in the $q$-state
Potts model has a simple dependence on $q$: it is exactly equal to the
density of particles in the coalescence model, \coal, times $(q-1)/q$.

For completeness, we note that for low initial concentrations of kinks,
passage to the continuum limit is justified.  Putting 
$x=na$, $G_n(t)\to G(x,t)$, and $\Gamma=D/a^2$ in eq.~(3), and taking the
limit $a\to0$, one obtains the diffusion equation
$$
{\partial\over\partial t}G(x,t)=2D{\partial^2\over\partial x^2}G(x,t)\;.
  \eqno(10)
$$
The boundary condition~(4a) becomes $G(0,t)=1$, and the concentration
is obtained from the continuum limit of eq.~(1); $c(t)=-\partial G/\partial
x|_{x=0}$.  Consider, for example, an initial concentration $c_0$ of randomly
placed kinks, where the states of the domains in between are random (but do
not repeat).  Following simple combinatorial arguments, it can be shown that
this corresponds to the initial condition
$$
G(x,0)={1\over q}+{q-1\over q}e^{-{q\over q-1}c_0x}\;.
$$
Solving for the concentration, we obtain
$$
c(t)=c_0e^{z^2}{\rm erfc}(z)\;,\qquad z={q\over q-1}c_0\sqrt{2Dt}\;,
$$
which has the expected long-time asymptotic behavior
$$
c(t)\sim{q-1\over q}{1\over\sqrt{2\pi Dt}}\;, \qquad t\to\infty\;.
$$

We have introduced a new method for the analysis of dynamic lattice models in
one dimension.  This approach brings to light the connection between
diffusion-limited coalescence, \coal, diffusion-limited annihilation, \annil,
and the $q$-state Potts model with Glauber dynamics in the zero-temperature
limit.  All three models are described by the {\it same\/} diffusion equation;
the reaction between particles (or kinks) is manifested in the boundary
condition~(4a), again, shared by all three models; and the only
differences arise from the different initial conditions implied in each case.

For random initial conditions we obtain results which are in agreement with
what is known from studies of persistence, but the new formalism can
also handle correlated initial conditions, where the three models might differ
in nontrivial ways.  Moreover, the new approach can be extended to
inhomogeneous situations, along the same lines as the method of empty
intervals (IPDF)~[14].

We thank L. Glasser and D. Kessler for edifying discussions,
and we gratefully acknowledge NSF support of this work (PHY-9820569).

\end{multicols}
\end{document}